\documentclass[12pt]{iopart}

\usepackage{graphicx}
\usepackage{cite}
\usepackage{color}  
\usepackage{ulem}
\usepackage{tabularx}

\begin{document}

\title[Transport control of dust particles via the EAE]{Transport control of dust particles via the Electrical Asymmetry Effect: experiment, simulation, and modeling}

\author{Shinya Iwashita$^1$, Edmund Sch\"ungel$^1$, Julian Schulze$^1$, Peter Hartmann$^2$, Zolt\'an Donk\'o$^2$, Giichiro Uchida$^3$, Kazunori Koga$^3$, Masaharu Shiratani$^3$, Uwe Czarnetzki$^1$}

\address{$^1$ Institute for Plasma and Atomic Physics, Ruhr University Bochum, 44780 Bochum, Germany\\
$^2$ Institute for Solid State Physics and Optics, Wigner Research Centre for Physics, Hungarian Academy of Sciences, H-1525 Budapest POB 49, Hungary\\
$^3$ Department of Electronics, Kyushu University, 819-0395 Fukuoka, Japan}
\ead{shinya.iwashita@rub.de}

\begin{abstract}

The control of the spatial distribution of micrometer-sized dust particles in capacitively coupled radio frequency 
discharges is relevant for research and applications. Typically, dust particles in plasmas form a layer located at 
the sheath edge adjacent to the bottom electrode. Here, a method of manipulating this distribution by the 
application of a specific excitation waveform, i.e. two consecutive harmonics, is discussed. 
Tuning the phase angle $\theta$ between the two harmonics allows to adjust the discharge symmetry via the Electrical 
Asymmetry Effect (EAE). An adiabatic (continuous) phase shift leaves the dust particles at an 
equilibrium position close to the lower sheath edge. Their levitation can be correlated with the electric field 
profile. By applying an abrupt phase shift the dust 
particles are transported between both sheaths through the plasma bulk and partially reside at an equilibium 
position close to the upper sheath edge. Hence, the potential profile in the bulk region is probed by the dust 
particles providing indirect information on plasma properties. The respective motion is understood by an analytical model, 
showing both the limitations and possible ways of optimizing this sheath-to-sheath transport. A classification 
of the transport depending on the change in the dc self bias is provided, and the pressure dependence is discussed.

\end{abstract}

\maketitle

\section{Introduction}
\label{Introduction}
Dusty plasmas exhibit interesting physical phenomena \cite{DustyPlasmasBasic,FortovPysRep2005} such as the interaction of the plasma 
sheath \cite{ParticleSheath1,ParticleSheath2,ParticleSheath3,Melzer} and bulk \cite{ParticleBulk} with the dust particles, the 
occurrence of waves \cite{ParticleWaves1} and instabilities \cite{ParticleInstab1,ParticleInstab2,ParticleInstab3}, 
phase transitions \cite{Phase1,Phase2,Phase3,Phase4,Phase5}, and the formation of Coulomb crystals \cite{Thomas,Chu,Hayashi,Arp}. 
They have drawn a great attention for industrial application because dust particles in plasmas play 
various roles: on one hand the accumulation of dust particles is a major problem for device operation 
in fusion plasma reactors as well as for semiconductor manufacturing \cite{Bonitz,Shukla,Bouchoule,Krasheninnikov,Selwyn}, 
i.e. they are impurities to be removed. On the other hand, they are of general importance for deposition 
purposes \cite{DustDepo1,DustDepo2} and it is well known that an enhanced control of such dust particles in plasmas 
has the potential to realize the bottom up approach of fabricating novel materials, e.g., microelectronic circuits, 
medical components, and catalysts \cite{ShirataniJPD11,Koga,Wang,Yan,Fumagalli,Kim}. In all cases the manipulation 
of dust particles, which is realized by controlling forces exerted on them such as electrostatic, thermophoretic, 
ion drag, and gravitational forces, or externally applied ones, e.g., created by a laser 
beam \cite{Nosenkoa,MorfillPoP10,Laser1,Laser2}, is crucially important. Furthermore, the use of dust particles as 
probes of these forces revealing plasma properties is a current topic of research \cite{MorfillPRL04,DustProbes2}.\\
\indent We have developed a novel method to control the transport of dust particles in a capacitively 
coupled radio frequency (CCRF) discharge by controlling the electrical symmetry of the discharge \cite{Iwashita}. 
Alternative dust manipulation methods using electrical pulses applied to wires have also been 
reported \cite{SamsonovPRL2002,PustylnikPRE2006,KnapekPRL2007,PustylnikPoP2009}.
Our dust manipulation method is based on the Electrical Asymmetry Effect (EAE) \cite{Heil}. 
The EAE allows to generate and control a dc self bias, $\eta$, electrically even in geometrically symmetric 
discharges. It is based on driving one electrode with a particular voltage waveform, $\phi_{\sim}(t)$, which is the 
sum of two consecutive harmonics with an adjustable phase shift, $\theta$: 
\begin{equation}
\label{EQappvol}
\phi_{\sim}(t)=\frac{1}{2}\phi_0[\cos(2\pi f t+\theta)+\cos(4 \pi f t) ]. 
\end{equation}
Here, $\phi_0$ is the identical amplitude of both harmonics. In such discharges, $\eta$ is an almost linear function 
of $\theta$. In this way, separate control of the ion mean energy and flux at both electrodes is realized 
in an almost ideal way. At low pressures of a few Pa, the EAE additionally allows one to control the maximum 
sheath voltage and width at each electrode by adjusting $\theta$ \cite{Heil}, resulting in 
the control of forces exerted on dust particles, such as electrostatic and ion drag forces. 
In contrast to the pulsing methods mentioned above, the change in the phase angle does not 
require a change in the applied power or RF amplitude. Furthermore, it is a radio frequency technique, i.e. no DC voltage is 
applied externally and the EAE is, therefore, applicable to capacitive discharge applications with 
dielectric electrode surfaces, without the need for additional electrodes or power supplies for the pulsing. 
The EAE can be optimized with respect to the control range of the dc self-bias by choosing non-equal voltage 
amplitudes for the individual harmonics \cite{EAE7} or by adding more consecutive harmonics to the applied 
voltage waveform \cite{EAE11,BoothJPD12}. In this study we intend to describe the basic mechanisms of the 
manipulation of the dust particle distribution in electrically asymmetric CCRF discharges. Thus, we restrict 
ourselves to the simplest case described by Eq. (\ref{EQappvol}). 
It is important for the analysis carried out in this work that the dust density is sufficiently low so that the 
plasma parameters are not disturbed by the dust particles. 
A large concentration of dust particles disturbs the electron density and can cause a significant change of the dc self 
bias when distributed asymmetrically between the sheaths \cite{Boufendi2011,Watanabe1994,EddiJPD2013}. The critical 
parameter for the disturbance is Havnes' value: $P = 695 T_e r_d n_d / n_i$, where $T_e$, $r_d$, $n_d$ and $n_i$ are 
electron temperature, radius of dust particles, their number density and ion density, respectively \cite{Thomas,Havnes1990}. 
$P$ is basically the ratio of the charge density of dust particles to that of ions. The concentration of dust particles 
disturbs the electron density for $P > 1$, while it does not for $P << 1$. 
In the critical region ${P_c} = 0.1-1$ the charge of the dust particles becomes significant in the total charge 
balance \cite{Havnes1990}. We calculate $P \approx 10^{-3}$ for our experiment, which is well below the $P_c$. For 
this estimation, direct images of dust particles were analyzed and a mean distance between particles of about 1 mm is 
determined. Thus, the concentration of dust particles is quite low in this study and they do not disturb the plasma. \\
\indent This paper is structured in the following way: this introduction is followed by a description of the 
methods used in this work. There, information on the experimental 
setup as well as the numerical simulation method is provided, and 
the analytical approaches on the RF sheath driven by non-sinusoidal voltage 
waveforms and the motion of dust particles in the plasma bulk region are explained. 
The results, which are presented and discussed in the third 
section, include the control of the dc self bias in dusty plasmas via the EAE, the change of the dust levitation 
position when changing the phase angle adiabatically (continuously), the motion of dust particles through the plasma bulk 
when tuning the phase angle abruptly, and a classification of the dust particle transport depending on the change 
in the dc self bias and the discharge conditions. Finally, concluding remarks are given in section four. 

\section{Methods}

\subsection{Experiment}

\begin{figure}[b]
\begin{center}
\includegraphics{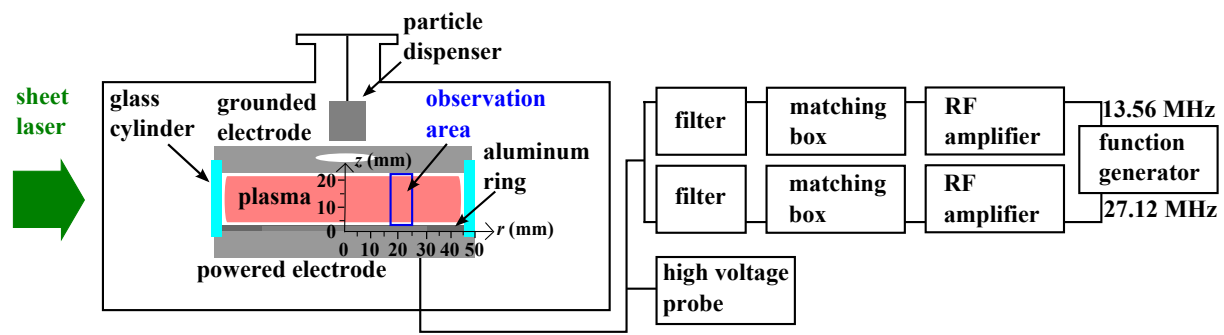}
\caption{Sketch of the experimental setup.}
\label{FIGsetup}
\end{center}
\end{figure}

Figure \ref{FIGsetup} shows the experimental setup. The experiments are carried out using a CCRF discharge operated
in argon gas at $p$ = 2 - 13 Pa, excited by applying $\phi_{\sim}(t)$ according to Eq. (\ref{EQappvol}) with $f$ = 
13.56 MHz and $\phi_0$ = 200 - 240 V. The applied voltage and the dc self bias are 
measured using a high voltage probe. Details of the electrical circuit have been provided in previous papers 
\cite{Julian,Iwashita}. The lower (powered) and upper (grounded) electrodes of 100 mm diameter are placed at a distance of 
$d=22$ mm. The plasma is confined radially between the electrodes by a glass cylinder to improve the discharge 
symmetry. Both the grounded chamber and the powered electrode are water cooled to 
eliminate the influence of the thermophoretic force. The upper electrode has a 20 mm diameter hole sealed with a 
fine sieve in the center for injecting SiO$_2$ dust particles of 1.5 $\mu$m in size, from a dispenser situated 
above the upper electrode. The gap between the upper electrode and the dispenser, which is located at the center 
of the upper electrode, is sealed with a teflon ring to prevent any disturbances due to gas flowing through the 
gap. The supply of argon gas inside the glass cylinder is realized through slits of a teflon ring, which is placed 
between the glass cylinder and the grounded electrode. An aluminum ring (100 mm outer diameter, 60 mm inner 
diameter, 2 mm height) is set on the lower electrode to confine the dust particles radially. The injected dust 
particles initially tend to reside relatively near the edge inside the aluminum ring, therefore 
the observation area is taken to be in the region of 2 mm $\leq z \leq$  22 mm and 18 mm $\leq r \leq $ 25 mm 
using a two dimensional laser light scattering 
(2DLLS) method \cite{Bouchoule,ShirataniJPD11,Koga,XuLLS} as shown in Fig. \ref{FIGsetup}. A vertical laser sheet 
passes between the two electrodes, with height and width of 20 mm and 1 mm, respectively. The laser power is 
150 mW at 532 nm. The light scattered by the dust particles is detected through a side window using a CCD 
camera equipped with an interference filter and running at a frame rate of 30 pictures per second. 

\subsection{PIC/MCC simulation}
\begin{figure}[t]
\begin{center}
\includegraphics{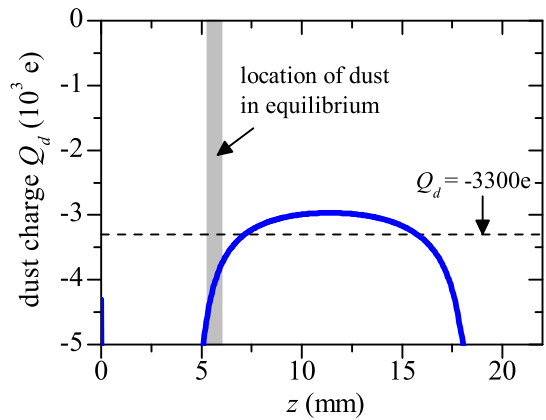}
\caption{Estimated spatial profile of the dust charge based on the standard formula \cite{Bonitz,Piel} (Ar, 8 Pa, $\phi_0$ = 200 V, $\theta$ = $0^{\circ}$). 
The dashed line shows the spatial average in the plasma bulk, that is used in the manuscript. The location of dust particles in equilibrium near the 
lower electrode, which is obtained experimentally, is also shown.}
\label{Dustcharge} 
\end{center}
\end{figure}

The rf discharge is described by a simulation code based on the Particle-In-Cell approach combined with Monte Carlo 
treatment of collision processes, PIC/MCC \cite{DonkoJPD09,DonkoAPL09,DonkoPSST11}. The code is one-dimensional in space and 
three-dimensional in velocity space. The simulations are performed in pure argon, although PIC/MCC simulations of dusty plasmas 
have already been reported \cite{Choi,Schweigert,Matyash}. Our approximation is based on the assumption that the dust particles 
represent only a minor perturbation to the plasma, which is justified for low concentration of dust particles as it is the case in this study. 
It has been proven that the simulations can be used to explain the motion of dust particles qualitatively as 
described in \cite{Iwashita}, and the forthcoming analysis also shows the applicability. 
The PIC/MCC simulations are performed at pressures between 4 and 12 Pa. 
Although our simulations are not capable of accounting for any two dimensional effects, the simulation data are 
helpful to understand the experimental findings, which are analyzed in the direction perpendicular to the electrode 
surfaces only. In the simulations the discharge is driven by a voltage specified by Eq. (\ref{EQappvol}). 
Electrons are reflected from the electrode surfaces with a probability of 0.2 and the secondary electron emission 
coefficient is set to $\gamma$ = 0.1. Based on the simulation results, the time averaged forces acting on dust 
particles, i.e. the ion drag force, $F_i$, electrostatic force, $F_e$ and gravity, $F_g$, are calculated as a function 
of the position between the electrodes \cite{Iwashita}. Here, the model of $F_i$ provided by Barnes et al \cite{Barnes} 
is applied. 
$F_e$ and $F_g$ are simply expressed as $F_e = Q_dE$ and 
$F_g = m_dg$, where $E$ and $m_d$ are the time averaged electric field and mass of dust particles, respectively. 
The charge of dust particles is calculated based on the standard formula: $Q_d = 1400 r_d T_e$ for isolated dust 
particles, e.g., by Bonitz \cite{Bonitz} or Piel \cite{Piel}, to be $Q_d \approx -3300e$ in the plasma bulk 
(see Fig. \ref{Dustcharge}), which is close to the typical value reported elsewhere \cite{ParticleBulk}. 
Here $e$ is the elementary charge. 
The typical error in the plasma bulk due to the spatial inhomogeneity is estimated to be about 10 \%. Finally, 
the spatial profiles of the potential energy are derived from the net forces exerted on dust particles.

\subsection{Analytical model of the RF sheath driven by an arbitrary voltage waveform}
\label{SectionSheathModel}
In this section a model of CCRF discharges is combined with the Child-Langmuir approximation to obtain the 
main properties of the RF sheath, i.e. the time dependent sheath width and the spatio-temporal distribution of the 
potential and electric field inside the sheath, in an electrically asymmetric capacitive discharge. 
The goal is to calculate the time averaged sheath electric field and correlate this field with the 
levitation of the dust particles above the powered electrode in case of an adiabatic phase shift, discussed in 
section \ref{adiabatic}.
The dynamics of the sheath in a "classical" dual frequency discharge driven by two substantially different frequencies has been 
modeled using similar approaches \cite{DFsheath1,DFsheath2,DFsheath3}.
According to the model, which has been introduced in \cite{Heil,VQ,Czarnetzki11}, we find the following expression for 
the sheath voltage at the powered electrode normalized by $\phi_0$: 
\begin{equation}
\bar{\phi}_{sp}(t) = - \left[\frac{-{\varepsilon}q_t + \sqrt{{\varepsilon}{q_t}^2 - (1-\varepsilon)[\bar\eta + \bar\phi_{\sim}(t)]}}{1 - \varepsilon}\right]^2.
\label{EQphibar}
\end{equation}
Here $\varepsilon$, $q_t$, $\bar\eta$ and $\bar\phi_{\sim}(t)$ are the symmetry parameter as defined and discussed in \cite{Heil}, 
normalized total charge, the dc self bias as well as the applied voltage normalized by $\phi_0$, respectively. 
Eq. (\ref{EQphibar}) provides the sheath voltage as a function of time. In order to obtain a spatio-temporal 
model of the sheath electric field, the collisionless Child-Langmuir sheath theory \cite{Lieberman} can be applied 
at low pressures of a few Pa.
To simplify the analysis, we restrict ourselves to a one-dimensional scenario. In this approximation, 
the maximum width of the sheath adjacent to the powered electrode is expressed as 
$s_{max,p} = \frac{\sqrt{2}}{3} \lambda_{De} \left(2 \left| \hat{\phi}_{sp} \right| e / T_e\right)^\frac{3}{4}$, 
where $\hat{\phi}_{sp}$, $\lambda_{De}$ and $T_e$ are the maximum of the sheath voltage at the powered electrode, 
the Debye length and the electron temperature at the sheath edge (in eV), 
respectively. 
The time dependent sheath width is given by the scaling with the sheath voltage:
$s_p(t) = s_{max,p} \left( \phi_{sp}(t) / \hat{\phi}_{sp} \right)^{\frac{3}{4}}$. 
The minimum voltage drop across the powered sheath, $\hat{\phi}_{sp} <0$, is found from the voltage balance: 
$\phi_{\sim}(t) + \eta = \phi_{sp} + \phi_{sg} + \phi_b$ 
at the time of minimum applied voltage. 
Here $\phi_{sg}$ and $\phi_b$ are the sheath voltage at the grounded electrode and the bulk voltage, respectively. 
Neglecting the floating potential at the 
grounded sheath and $\phi_b$ yields $\hat{\phi}_{sp} \approx \tilde\phi_{min}+\eta,$
so that the minimum sheath voltage can easily be deduced from experimentally measured values, for instance. 
Here $\tilde\phi_{min}$ is the minimum of the applied voltage. 
Assuming 
that both the electric field and the potential are zero at the sheath edge the spatio-temporal profile 
of the electric potential in the sheath region at the powered electrode ($0 \leq z \leq s_p(t)$) is expressed by \cite{Oksuz} 
\begin{equation}
\label{EQsheathvoltage}
\phi_{sp}(z,t) = - \frac{T_e}{2e} \left(  \frac{3}{\sqrt{2}}\frac{s_p(t)-z}{\lambda_{De}}  \right)^{\frac{4}{3}}.
\end{equation}
Here $z$ = 0 is the position of the powered electrode. 
Finally, the spatio-temporal profile of the electric field in the sheath region is found by differentiation:
\begin{equation}
\label{Efield}                                                                         
E_{sp}(z,t)=-\frac{\partial \phi_{sp}(z)}{\partial z}=-\frac{\sqrt{2} T_e}{e \lambda_{De}}\left( \frac{3}{\sqrt{2}}\frac{s_p(t)-z}{\lambda_{De}} \right)^{\frac{1}{3}}
\end{equation}
Eq. (\ref{Efield}) is used to understand the dust motion as a consequence of the adiabatic (continuous) phase change and to determine the electron density 
in section \ref{adiabatic}.  

\subsection{Model to describe the motion of dust particles}
\label{SectionTransportModel}
\begin{figure}[b]
\begin{center}
\includegraphics{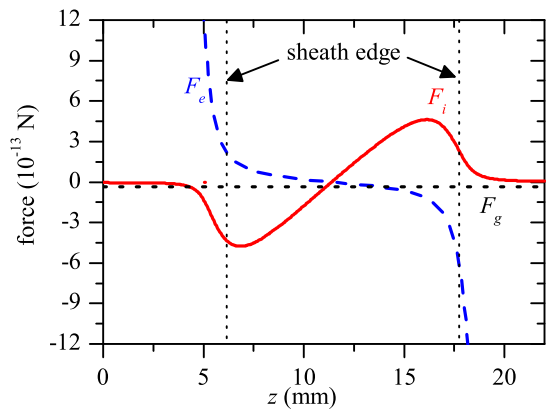}
\caption{Spatial profile of electrostatic force, $F_e$, ion drag force, $F_i$, and gravity, $F_g$, 
exerted on dust particles. The spatial profile is obtained from PIC/MCC simulation 
(Ar, 8 Pa, $\phi_0$ = 200 V, $\theta$ = $0^{\circ}$).} 
\label{Forces} 
\end{center}
\end{figure}
The motion of dust particles in plasmas is determinded by the forces exerted on them 
\cite{Bouchoule,Barnes,Garrity,Morfill,Goree,Piel}. Here, we propose a simple analytical model to describe 
the one-dimensional transport of dust particles between both sheaths through the plasma bulk. 
Models of the dust motion based on the force balance have already been reported \cite{Chu,Couedel, NefedovNJP2003,LandNJP2007,Zhakhovskii,Graves}. 
We would like to emphasize again that the concentration of dust particles is quite low in this study and they do not disturb the plasma, which 
is different from the condition under which these models have been provided. 
Our approach focuses on analyzing the particular dust transport which has been obtained experimentally 
when changing the phase angle abruptly, and in fact the model proposed here can explain the experimental results. 

\begin{figure}[b]
\begin{center}
\includegraphics{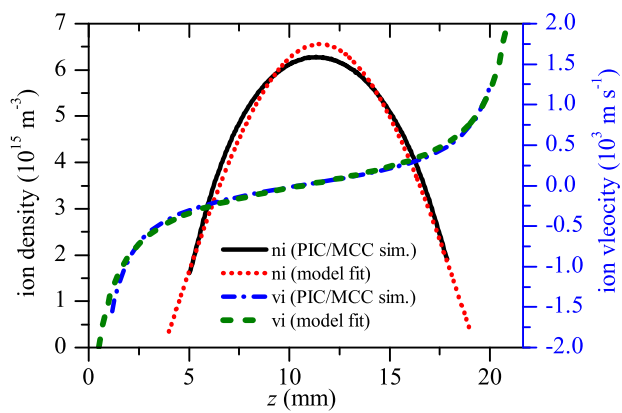}
\caption{Spatial profile of ion density and velocity obtained from the PIC/MCC simulation and fit functions of the analytical model 
(Ar, 8 Pa, $\phi_0$ = 200 V, $\theta$ = $0^{\circ}$).} 
\label{Fitting} 
\end{center}
\end{figure}

Further studies are required to investigate non-Hamiltonian effects \cite{Tuckerman,Kompaneets} and 
clarify their role for the physics presented in this work. f
In reactors with horizontal plane parallel electrodes separated by a discharge gap, $d$, and in the absence of thermophoretic 
forces, negatively charged dust particles tend to be confined at the sheath edges, where the forces exerted 
on them balance. Right after introducing the dust particles into the discharge volume, they are typically 
located around the lower sheath edge due to gravity. Let us focus on the motion of dust particles between the 
sheath edge of the bottom (powered) electrode $(z = s_p)$ and the upper (grounded) one $(z = d-s_g)$, e.g., 
after applying an upward force at the lower equilibium position. Later on, we will approximate the electrostatic 
force around the sheath edges as hard walls, i.e. the particles are instantaneously reflected without any change 
in their kinetic energy. 
This assumption is justified due to the fact that the electrostatic force caused by the bulk electric field 
(see Fig.  \ref{Forces}) or the interaction between dust particles is negligible under our condition. 
One reason for this quite small bulk electrostatic force is the relatively high ion density in the bulk, which is also 
realized in the void formation in dusty plasmas \cite{Bouchoule,Bonitz}.
In contrast to our situation, the electrostatic force is of vital importance in complex plasmas, where the major contribution of 
negative charges to the total charge balance in the bulk is given by the dust particles and not by the electrons (see e.g., \cite{Chu,Couedel,Takahashi}). 
The inter-particle force, i.e. Coulomb force can be comparable to the sheath electrostatic force under certain conditions \cite{Hwang}. 
This becomes crucial particularly when the lateral motion of dust particles is discussed. This study is, however, focused only on 
their vertical motion. Additionally, dust particles are initially located only at the lower sheath edge due to the balance between 
the sheath electrostatic force and the ion drag force, suggesting that these two forces are dominantly exerted on the dust particles in this study. 
Thus, the vertical component of the Coulomb force is much smaller than the respective component of the sheath electrostatic force and the ion drag force. 
In our model, 
small errors occur only at the bulk side of the sheath edge (equilibrium position of dust particles) where the electrostatic force 
is neither close to zero nor represents a hard wall. The dust particles are assumed not to perturb the plasma. Within the plasma 
bulk region, the dust particle motion is associated with the following force balance:
\begin{equation}
\label{EQmombal}
m_d\ddot{z} = -m_dg - m_d\nu\dot{z} + F_i(z).
\end{equation}
Here, $m_d$, $g$, $\nu$, and $F_i$ are the mass of a dust particle, the acceleration of gravity, the frequency 
of momentum loss due to collisions between dust particles and gas atoms \cite{Piel,Epstein}, and the ion drag 
force, respectively. Note that the gas friction force $m_d\nu\dot{z}$ is derived from the assumption that the 
velocity of dust particles is much smaller than the thermal velocity of gas molecules. Therefore, the dependence 
of $\nu$ on the particle velocity can be neglected. Any interaction between the dust particles, e.g., a repulsive Coulomb 
force \cite{Thomas,Chu,Hayashi,Arp,Takahashi,LinIJPD94}, is not taken into account.
Although the force profiles shown in Fig. \ref{Forces} suggest that gravity can be neglected, we keep the corresponding term 
in the force balance to ensure the applicability of the resulting formulae for all types of particles, e.g., different sizes and/or mass densities (materials).\\
\indent There are several models of the ion drag force \cite{Barnes,KhrapakPRL2003,Fortov2} and the analytical 
description of this force remains an interesting research topic in itself. There are discussions in the literature on the validity 
of the different models. Although more sophisticated models are available, the Barnes model \cite{Barnes} is applied here in order 
to calculate the ion drag force in a simple way. The formula is generally considered to be accurate at low dust densities as pointed out 
e.g., in \cite{Bouchoule,Piel}, which is the case in this study. 
We assume that $n_i$ as well as the ion velocity, $v_i$, are expressed by trigonometric functions, 
as it results from the basic diffusion estimation in a steady state CCRF discharge \cite{Lieberman}: 
\begin{eqnarray}
\label{iondensityprofile}
n_i (z) & = & n_{i0} \cos\left[\left(z-\frac{d}{2}\right)\frac{\pi}{\Lambda_{i}}\right], \\
v_i (z) & = & v_{i0} \tan\left[\left(z-\frac{d}{2}\right)\frac{\pi}{\Lambda_{i}}\right].
\label{ionvelocityprofile}
\end{eqnarray}
\noindent Here, the maximum ion density, $n_{i0}$, and ion velocity, $v_{i0}$, are constants. $\Lambda_{i}$ is the ion diffusion length; 
the value is actually close to the distance between the discharge center and the sheath edges. These input parameters are 
determined by fitting to the PIC/MCC simulation data as shown in  Fig. \ref{Fitting}. 
The estimated model quantities from this fitting are $n_{i0}=6.6 \cdot 10^{15}$ m$^{-3}$, $v_{i0}=344$ m s$^{-1}$, $\Lambda_i=15.5$ mm, 
and $d=23.0$ mm, respectively. 
The ion drag force consists of the collection force due to ions hitting the particle surface and the orbit force due to Coulomb collisions with 
the drifting ions. In low pressure CCRF discharges the orbit force \cite{Barnes}, 
\begin{equation}
F_{i,orb}= 4 \pi n_i v_s m_i b_{\pi/2}^2 \Gamma,
\end{equation}
typically dominates. Here, $v_s$, $m_i$, $b_{\pi/2}$ and $\Gamma$ are the mean ion velocity, the ion mass, 
the impact parameter and the Coulomb logarithm \cite{Barnes}, respectively:
\begin{eqnarray}
b_{\pi/2} & = & \frac{eQ_d}{4\pi\epsilon_0m_iv_s^2}, \\
\Gamma & =& \frac{1}{2}\ln\left(\frac{\lambda_{De}^2+b_{\pi/2}^2}{r_d^2(1-\frac{2e\phi_f}{m_i v_s^2})+b_{\pi/2}^2}\right).
\end{eqnarray}
Note that these quantities depend on the radius ($r_d$), floating potential ($\phi_f$), and charge ($Q_d$) of the 
dust particles. In this paper, we use the simplifying assumption of the dust particle charge to be negative and 
constant: $Q_d \approx -3300e$ as shown in Fig. \ref{Dustcharge}. \\
\indent In our approach, we neglect the thermal motion of the ions, i.e. the mean ion velocity $v_s$ is given by 
the drift component, $v_i$:
\begin{equation}
v_s=\left(\frac{8k_BT_i}{\pi m_i}+v_i^2\right)^{\frac{1}{2}} \approx v_i.
\end{equation}
Applying the approximation $F_i \approx F_{i,orb} \propto n_iv_i$, the ion drag force becomes
\begin{equation}
F_i (z) = \bar{F}_{i0} \sin\left[\left(z-\frac{d}{2}\right)\frac{\pi}{\Lambda_i}\right].
\end{equation}
\noindent Here, the maximum ion drag force ($\bar{F}_{i0}$) is a constant. In order to solve Eq. (\ref{EQmombal}) 
analytically only the linear variation of the sine function is considered here: 
\begin{equation}
\label{EQiondrag}
F_i (z) \approx F_{i0} \left(z-\frac{d}{2}\right) \frac{\pi}{\Lambda_{i}},
\end{equation}
with $F_{i0}= 4 \pi m_i n_{i0} v_{i0} b_{\pi/2}^2 \Gamma$. 
The input parameters obtained from Fig. \ref{Fitting} provide $F_{i0}=3.8 \cdot 10^{-13}$ N. 
Equation \ref{EQiondrag} corresponds to a strong simplification of 
$F_i (z)$ and deviations from the exact solution appear, particularly in the 
regions close to the sheath edges. However, our aim is to explain the transport of dust particles through 
the plasma bulk with this model. In the bulk region, the model is a reasonable approach, since it includes the 
most relevant forces in this region. Furthermore, the forthcoming analysis shows that the basic features of particle motion 
and the experimental observation of the dust transport can be explained reasonably well by this approach. \\
\indent After inserting Eq. (\ref{EQiondrag}) into Eq. (\ref{EQmombal}) a second order 
linear ordinary differential equation 
\begin{equation}
\label{EQmombalsimple}
m_d\ddot{z} + m_d\nu\dot{z} - F_{i0} \left[\left(z-\frac{d}{2}\right)\frac{\pi}{\Lambda_{i}}\right] + m_dg = 0
\end{equation}
\noindent needs to be solved. Note that Eq. (\ref{EQmombalsimple}) represents a harmonic oscillator 
in the space coordinate $(z-d/2)\pi/\Lambda_{i}$ with frequency $\sqrt{F_{i0}/m_d}$, which is externally driven by gravity 
and damped by collisions. Finally, using the boundary conditions $z(0)=z_0$ and $\dot{z}(0)=u_{0}$, 
which corresponds to the initial velocity of dust particles, the trajectory of dust particles is given by
\begin{equation}
\label{EQtrajectory}
z(t) = \left[\beta_1 \cosh \left(\alpha t \right) + \beta_2 \sinh \left(\alpha t\right)\right] e^{-\frac{\nu}{2}t} + \delta.
\end{equation}
\noindent Here, $\alpha$, $\beta_1$, $\beta_2$, and $\delta$ are:
\begin{eqnarray}
\alpha & = & \sqrt{\left( \frac{\nu}{2}\right)^2 + \frac{\pi F_{i0}}{m_d \Lambda_{i}}}, \\
\beta_1 & = & x_0 - \frac{d}{2} - \frac{m_d \Lambda_{i} g}{\pi F_{i0}}, \\
\beta_2 & = & \left( u_0 + \beta_1 \frac{\nu}{2} \right) \alpha^{-1}, \\
\delta & = & \frac{m_d \Lambda_{i} g}{\pi F_{i0}} + \frac{d}{2}. 
\end{eqnarray}
From this trajectory of the dust particles, the kinetic energy is obtained:
\begin{equation}
\label{EQenergy}
W(t) = \frac{1}{2} m_d\dot{z}^2(t) = \frac{m_d}{8\alpha^2} \left( -A e^{\alpha t} + B e^{-\alpha t} \right)^2 e^{-\nu t}, 
\end{equation}
\noindent where A and B are defined as
\begin{eqnarray}
A & = & g + \frac{F_{i0} \pi d}{2 m_d \Lambda_{i}} - x_0 \frac{F_{i0} \pi}{m_d \Lambda_{i}} + u_0 \left( \frac{\nu}{2} - \alpha \right), \\
B & = & g + \frac{F_{i0} \pi d}{2 m_d \Lambda_{i}} - x_0 \frac{F_{i0} \pi}{m_d \Lambda_{i}} + u_0 \left( \frac{\nu}{2} + \alpha \right).
\end{eqnarray}	
\noindent Eq. (\ref{EQenergy}) is used to describe the dust energy as a consequence of the abrupt phase change in section \ref{abrupt}. 
This rather complex result will be compared to the simple assumption that 
the kinetic energy of the dust particles is not affected by the particular shape of the potential profile and that the 
loss of the energy of the dust particles is only due to gas friction. 
Then, the velocity and kinetic energy of 
the dust particles can be estimated as 
\begin{eqnarray}
u_{d}(t) = u_{0}e^{-\frac{\nu}{2}t}, \\
\label{EQenergysimple0}
W(t) = \frac{1}{2}m_{d}u_{d}^{2}(t) = W_0e^{-{\nu}t} .
\label{EQenergysimple}
\end{eqnarray}
Here $W_0$ is the initial kinetic energy of dust particles. 
Eq. (\ref{EQenergysimple}) is used to determine the potential profile experimentally using the spatial 
profile of the laser light scattering (LLS) 
intensity from dust particles in section \ref{abrupt}. 
It should be noted that practically the dust charge fluctuates and the reflection of the dust particles at the sheath edge 
is ``soft''. Again, our model aims to describe the dust transport observed in this study in a simple way, 
and thus the simple assumption, e.g., a constant dust charge and a rough approximation of the electrostatic force as a hard wall, is applied here. 

\section{Results and Discussion}

\subsection{dc self bias control via the EAE in a plasma containing a small amount of dust}

\begin{figure}[b]
\begin{center}
\includegraphics{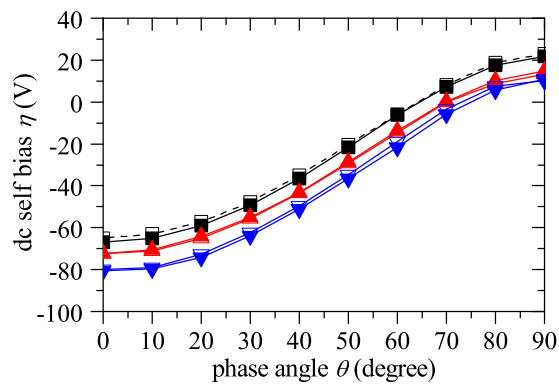}
\caption{Experimentally obtained dc self bias as a function of the phase angle $\theta$ with and without dust particles for 
different neutral gas pressures. The applied voltage amplitude is kept constant at $\phi_0$ = 200 V. 
Solid symbols relate to discharges without and open symbols to ones with dust particles. Square: 2 Pa, triangle: 4 Pa, 
inverted triangle: 8 Pa.}
\label{FIGbias} 
\end{center}
\end{figure}
 
Fig. \ref{FIGbias} shows the dc self bias, $\eta$, obtained from the experiment, as a function of the phase angle, 
$\theta$. $\eta$ is generated as a monotonic function of $\theta$. As described in details before 
\cite{Heil,EAE7,EAE11,Julian,Czarnetzki11,Eddi,DonkoJPD09,DonkoAPL09}, the EAE allows to control the discharge 
symmetry electrically. The control range 
for gas pressures between 2 and 8 Pa and an applied voltage amplitude of $\phi_0$ = 200 V 
is found to be close to about 45 \% of the applied voltage 
amplitude. Therefore, a strong change in both the time averaged sheath voltages 
($\eta=\left\langle \phi_{sp}(t)\right\rangle+\left\langle \phi_{sg}(t)\right\rangle$) and the maximum 
sheath voltages 
as a function of $\theta$ can be expected. $\eta$ is shifted 
towards negative values because the discharge setup becomes effectively geometrically asymmetric due to the 
parasitic effect of capacitive coupling between the glass cylinder and the grounded chamber walls 
\cite{Coburn,Savas,Julian,Booth10,Booth12,Booth12_2}. This effect tends to be stronger at higher pressures. 
It is important to note that in this study no significant difference of $\eta$ in cases with and without 
dust particles is observed, indicating that the presence of a low dust concentration does not influence the 
plasma significantly. Therefore, the models described in the previous section are indeed applicable 
as pointed out already in section \ref{Introduction} by estimating Havnes' value $P$.

\subsection{Adiabatic phase change}
\label{adiabatic}
\begin{figure}[b]
\begin{center}
\includegraphics{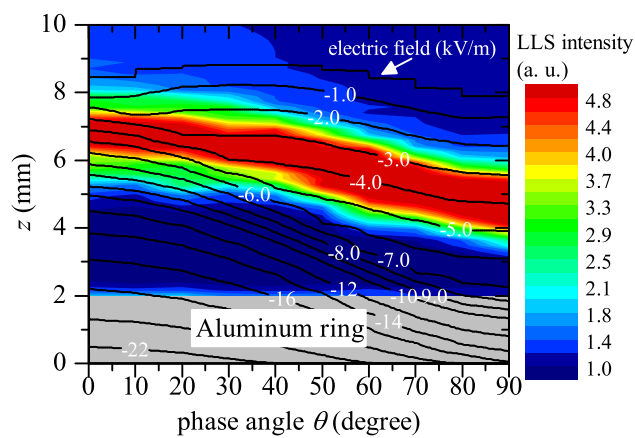}
\caption{Spatial profile of the measured LLS intensity from the dust particles around the lower electrode as a 
function of the phase angle $\theta$ combined with the electric field calculated from the analytical model 
(Ar, 2 Pa, $\phi_0$ = 200 V). The observation of the LLS intensity within the lower region (0 mm $\leq z \leq$ 2 mm) 
is blocked by the aluminum ring.}
\label{FIGllsfield}
\end{center}
\end{figure}
The dust particles injected into the discharge are initially located at the sheath edge adjacent to the lower 
electrode. Any adiabatic (continuous) change of $\theta$ leaves the dust particles at an equilibrium position 
close to this lower sheath edge as shown in Fig. \ref{FIGllsfield}. By increasing the phase angle from 
$0^{\circ}$ to $90^{\circ}$ adiabatically, the time averaged sheath width becomes 
smaller and both the mean and the maximum sheath voltages at the lower electrode decrease. Therefore, the 
equilibrium position of the dust particles is shifted closer towards the electrode. This change of the equilibrium 
position can be understood by the electric field profile obtained from the analytical model described in 
section \ref{SectionSheathModel} using input parameters of 
$T_e$ = 3 eV and $\lambda_{De}$ = 644 $\mu$m calculated 
under the assumption of $n_e$ = $4 \times 10^{14}$ m$^{-3}$ 
(see lines in Fig. \ref{FIGllsfield}). 
Electron density and temperature are taken from the PIC/MCC simulations 
because we applied a glass cylinder to confine the plasma. Thus, performing Langmuir probe measurements is not possible. 
We find very good agreement between the 
measured LLS and the part of the electric field distribution at a strength of about -4 kV/m, i.e. where 
forces exerted on dust particles balance. 
\begin{figure}[t]
\begin{center}
\includegraphics{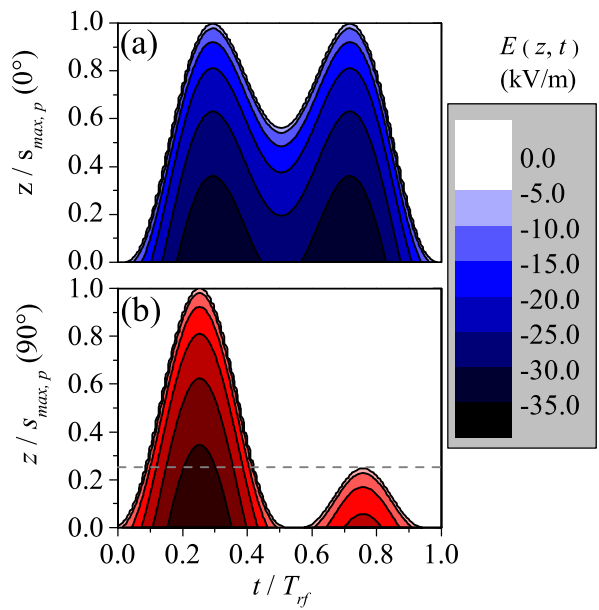}
\caption{Distribution of the electric field at (a) $\theta=0^\circ$ and (b) $\theta=90^\circ$ as a function of 
spatial position within the phase angle dependent maximum sheath width and time resulting from the model shown in 
Eq. (\ref{Efield}) (Ar, 2 Pa, $\phi_0$ = 200 V). $T_{rf}$ = 74 ns. 
The sheath reaches the region above dashed line only once per rf period.}
\label{FIGfield00field90}
\end{center}
\end{figure}
\begin{figure}[t]
\begin{center}
\includegraphics{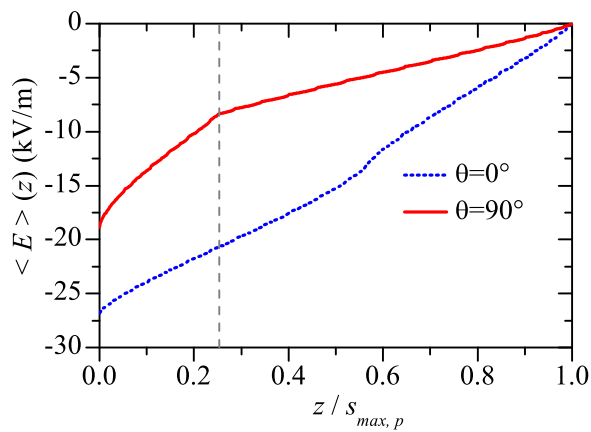}
\caption{Strength of the time averaged electric field as a function of position corresponding to Fig. \ref{FIGfield00field90}. 
The dashed line is drawn according to that in Fig. \ref{FIGfield00field90} (b). The gradient of the 
time averaged electric field changes at the boundary indicated by the dashed line.}
\label{FIGfield00field90_2}
\end{center}
\end{figure}
When $\theta$ is changed from $0^{\circ}$ to $90^{\circ}$, the maximum of the time averaged electric field 
in the powered electrode sheath, i.e. $\left\langle E\right\rangle_{max}$ found at the electrode, becomes 
smaller due to the decrease in the mean sheath voltage. In addition, the change in the shape 
of the applied voltage as a function of $\theta$ leads to a change in the sheath voltage, $\phi_{sp}(t)$, which causes a 
change in the spatial distribution of the time averaged electric field. As it becomes clear 
from Fig. \ref{FIGfield00field90} and \ref{FIGfield00field90_2}, the slope of $\left\langle E\right\rangle(z)$ becomes 
flatter in the upper part of the sheath with increasing $\theta$, i.e. the time averaged voltage drop over this region 
becomes smaller. 
In particular, the field is relatively small during the second half of the rf 
period (see dashed line in Fig. \ref{FIGfield00field90} (b)). 
Thus, the broadening of the equilibrium position (region of bright LLS) is well understood by the analytical model. This 
correlation analysis of the dust equilibrium position combined with the spatial electric field profile is 
applicable as a diagnostic tool to estimate plasma parameters, i.e. the dust particles can serve as electrostatic 
probes \cite{MorfillPRL04,Kersten09,DustProbes2,DustProbes3,DustProbes4}. The correlation analysis yields 
the maximum sheath extension as the only free fitting parameter, which depends on electron temperature and density 
($s_{max,p} \propto \lambda_{De} / T_e^{3/4} \propto n_e^{-1/2} T_e^{-1/4}$). 
Hence, $s_{max,p}$ is more sensitive to changes in the electron density and, if the electron temperature is 
known, $n_e$ can be obtained assuming that these plasma parameters are constant, independently of $\theta$. 
In our discharge configuration, it is not possible to measure $T_e$. However, estimating $T_e \approx 3$ eV, 
for instance, results in an electron density of about $n_e \approx 4 \cdot 10^{14}$ $m^{-3}$ at the sheath edge 
under the condition of Fig. \ref{FIGllsfield} (Ar, 2 Pa and $\phi_0$ = 200 V). Note that 
the charge of dust particles becomes smaller than that in the plasma bulk when they are closer to the sheath edge as shown in Fig. \ref{Dustcharge}, 
i.e. the charge of the dust particles observed 
in Fig. \ref{FIGfield00field90_2} might be smaller than -3300e which is assumed as the dust charge in this paper. 
Further study is required to discuss this topic in detail. 

\subsection{Abrupt phase change}
\label{abrupt}
\begin{figure}[t]
\begin{center}
\includegraphics{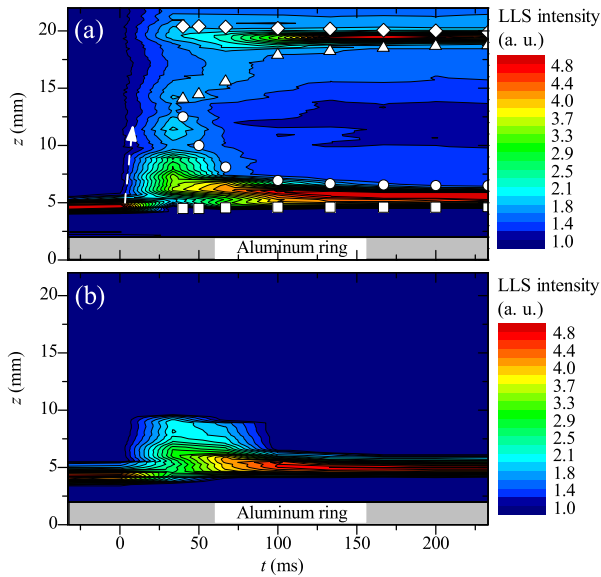}
\caption{Spatiotemporal profiles of the measured LLS intensity by the dust particles within the discharge gap 
(Ar, (a) 8 Pa and (b) 12 Pa, $\phi_0$ = 200 V). 
The abrupt phase change takes place at {\it t $\approx$} 0 ms. 
Observation of the lower region (0 mm $\leq z \leq$ 2 mm) is blocked by the aluminum ring. The upper 
(diamond and triangle) and lower (circle and square) points are taken to obtain the upper and lower 
potential wells in Fig. \ref{FIGpotexp}, respectively. The arrow illustrates the estimation of an initail 
velocity of $u_0 \approx 1$ m/s.}
\label{FIGtransport}
\end{center}
\end{figure}

\begin{figure}[ht]
\begin{center}
\includegraphics{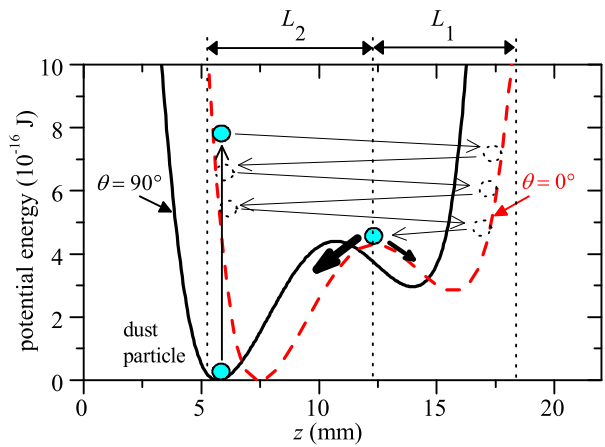}
\caption{Model of sheath-to-sheath transport of dust particles\cite{Iwashita}. 
The potential profile is calculated from PIC/MCC simulation data (Ar, 4 Pa, $\phi_0$ = 200 V). 
$L_1$ and $L_2$ are the widths of the upper and lower potential wells, respectively, at $\theta = 0^\circ$.}
\label{Transportmodel}
\end{center}
\end{figure}

\begin{figure}[t]
\begin{center}
\includegraphics{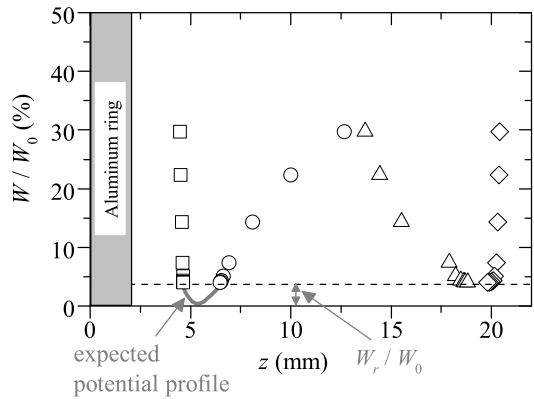}
\caption{Potential profile at $\theta = 0^\circ$ obtained from the measured 2DLLS intensity shown in 
Fig. \ref{FIGtransport} (a) using a simple model. The potential energy scale is normalized by a 
rough estimation of the initial energy of the dust particles.}
\label{FIGpotexp}
\end{center}
\end{figure}

\begin{figure}[t]
\begin{center}
\includegraphics{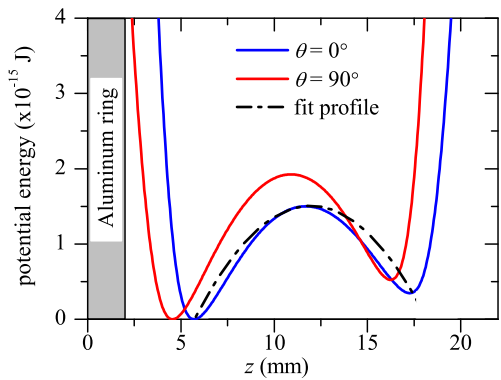}
\caption{Potential profile calculated from PIC/MCC simulations data (Ar, 8 Pa, $\phi_0$=200 V). 
The model curve resulting from fits of equations \ref{iondensityprofile} and \ref{ionvelocityprofile} to PIC simulation data is shown, as well.} 
\label{FIGpotpic}
\end{center}
\end{figure}

When the phase angle is changed abruptly from $90^{\circ}$ to $0^{\circ}$, i.e. much faster than the reaction 
time scale of the particles, all dust particles are transported upwards into the plasma bulk and undergo rapid 
oscillations between the sheaths. Thereafter, a fraction of the particles reaches the upper sheath region and 
settles there (see Fig. \ref{FIGtransport}(a)). In this way, sheath-to-sheath transport is realized \cite{Iwashita}. 
Before discussing the conditions, under which sheath-to-sheath transport is possible, in more detail, 
this particle motion should be understood. As in the case of the adiabatic phase change, dust particles injected into the 
discharge are initially located at the sheath edge adjacent to the lower electrode. If the phase is changed 
abruptly from $90^{\circ}$ to $0^{\circ}$, the dust particles are suddenly located in a region of high potential 
due to their inertia. Consequently, they bounce back and forth between both sheaths, while being decelerated by gas 
friction (see Fig. \ref{Transportmodel}) \cite{Iwashita}. 
As described in section \ref{SectionTransportModel}, 
the motion of dust particles is determined by gravity, the ion drag force pushing the particles out of the bulk 
towards the sheaths, deceleration due to friction by collisions with the neutral gas, as well as electrostatic 
forces due to the sheath electric field, which basically can be regarded as boundaries, thus spatially confining 
the particle motion. Afterwards, they reside inside the potential well at either the upper or the lower sheath 
edge \cite{Iwashita}. The shape of the potential profile consists of a peak close to the discharge center, two 
minima located around the sheath edges and steep rises inside the sheaths. The difference in the height of the two 
minima is mainly caused by gravity in the absence of thermophoretic forces. The term ``potential'' is valid only, 
if the result does not depend on the particle velocity, i.e. if the time scale of the dust particle motion is the 
slowest of all time scales of interest here. This condition is fulfilled: for instance, the thermal motion of 
both the neutral and the ionized gas atoms is about two orders of magnitude faster compared to the dust particle 
motion (the maximum dust velocity estimated from the experimental results (Fig. \ref{FIGtransport}) is a few m/s at most). 
Therefore, the potential profile is provided independently from the dust velocity. \\
\indent It is possible to determine this potential distribution qualitatively from the 
experimental results. Hence, information on basic plasma properties might be achievable from this analysis. 
The shapes of the potential wells at the upper and lower sheath edges are obtained from the LLS profile 
(see four kinds of points in Fig. \ref{FIGtransport}(a)). The points are taken at the contour 
line, which is both existent in the entire plasma bulk region and shows a reasonably high intensity. Note that 
the resulting data points are close to the region of maximum gradient of the LLS intensity, as well. 
The upper (diamond and triangle) and lower (circle and square) points correspond to the confinement regions of 
dust particles in the potential wells at the upper and lower sheath edges, respectively. In order to deduce the 
potential distribution from them, the temporal evolution of the energy of the dust particles needs to be known. 
The simplest model of the dust motion is applied here, i.e. dust particles lose their kinetic energy only due to 
gas friction. Using this approximation allows an analytical treatment of $W(t)$ by using Eq. (\ref{EQenergysimple}). 
Using the data points shown in Fig. \ref{FIGtransport} (a) and replacing the time scale by the corresponding energy, 
the potential profile shown in Fig. \ref{FIGpotexp} is obtained. Here, the potential energy scale is normalized by 
the initial energy of the dust particles. An estimation yields $W_0 \approx m_d u_0^2 /2 \approx 1.8 \times 10^{-15}$ J 
(11 keV) for an initial velocity of $u_0 \approx 1$ m/s, which was obtained from the spatiotemporal 
profile of the LLS intensity by the dust particles (see arrow in Fig. \ref{FIGtransport}). 
Taking into account the uncertainty in $W_0$, 
we restrict ourselves to a qualitative discussion of the potential profile in this 
study. Comparing this profile to the one calculated from the simulation data shown in Fig. \ref{FIGpotpic}, we see that 
the position of the lower potential minimum agrees well between the experiment and the PIC simulation 
($z \approx 5.7$ mm). In the experiment the upper minimum is located at 18.6 mm, whereas the position in the 
simulation is 16.9 mm. This difference is probably caused by the effective geometrical asymmetry of the discharge in 
the experiment, which is also indicated by the self bias voltage, $\eta$ (see 8 Pa case in Fig. \ref{FIGbias}). In 
the PIC simulation the discharge is geometrically symmetric, thus yielding a symmetric dc self bias curve 
($\eta(\theta=0^\circ)=-\eta(\theta=90^\circ) \approx -52$ V) and a wider sheath compared to the experiment at 
the grounded side for all $\theta$. The lowest part of the potential curve resulting from the experimental data 
cannot be obtained by this approach (see the curve at around z = 5 mm in Fig. \ref{FIGpotexp}), since the residual 
spatial distribution is caused by the residual energy, $W_r$, of dust particles in equilibrium 
position due to thermal motion and Coulomb interaction, respectively, as well as the spatial resolution of the 
optical measurements (see the LLS intensity from dust particles after 100 ms in Fig. \ref{FIGtransport}), which 
are neglected in our simple model. Except for this region, the dust particles can be used as probes to determine 
the potential, which depends on plasma properties via 
$F_i (z)$ and $F_{e}(z)$, in a major part of the discharge region. The probability for the trapping of dust 
particles at the upper sheath, $P_{trans}$ might be roughly estimated by the width of the upper potential well 
divided by the sum of the widths of the lower and upper potential wells	, which is expressed as $L_1 / (L_1 + L_2)$, 
in the simple approximation made above (see Fig. \ref{Transportmodel}) \cite{Iwashita}. 
Here $L_1$ and $L_2$ are the widths of the 
upper and lower potential wells, respectively. The probability calculated this way is 
about 0.5 for the experiment for 8 Pa and $\phi_0$=200 V, which agrees well with that 
calculated for the simulation potential profile. \\
\indent Furthermore, the potential profile can be used to obtain input parameters for the analytical model of 
dust transport described in section \ref{SectionTransportModel}. For this model the potential profile in the 
plasma bulk is obtained by integrating Eq. (\ref{EQmombalsimple}). 
Due to the small-angle approximation for the ion drag force (Eq. (\ref{EQiondrag})) the potential profile is 
expressed by a simple parabola: 
$U(z)=U_0 - \left[F_{i,0} \left( z - d \right) \frac{z \pi}{2 \Lambda_i} - m_d g z \right]$, where 
$U_0$ is an integration constant. 
The model curve resulting from fits of equations \ref{iondensityprofile} 
and \ref{ionvelocityprofile} to PIC simulation data is shown in Fig. \ref{FIGpotpic}. 
One can find a difference of the central maxima of the potential profile obtained from PIC/MCC simulation for $\theta = 90^{\circ}$ and $0^{\circ}$. 
This is derived from the spatial profiles of the ion drag force (mainly orbit force), i.e. the direction of the ion drag force changes at the 
center of the plasma bulk \cite{Iwashita} and the gradient of the force profile for $\theta = 90^{\circ}$ in this region 
is steeper than that for $\theta = 0^{\circ}$, resulting in the difference of the central maxima for $\theta = 90^{\circ}$ and $0^{\circ}$. 
The model shows reasonable agreement with the potential profile using the 
exact values from the PIC/MCC simulation within the plasma bulk. As discussed above, deviations can be observed close to the sheath edges, e.g., 
due to the simplified treatment of the electrostatic force as a hard wall. 

\begin{figure}[b]
\begin{center}
\includegraphics{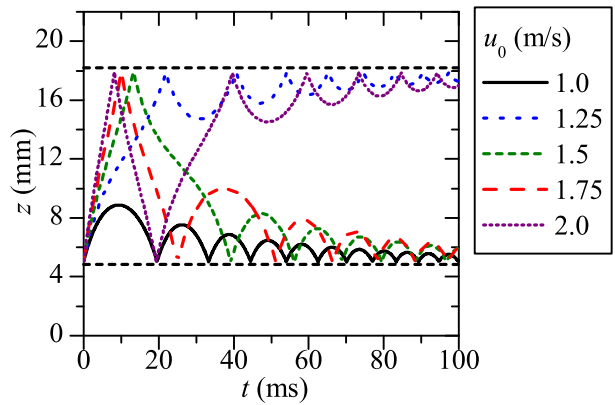}
\caption{Trajectory of dust particles calculated from the model for different initial velocities (Ar, 8 Pa, $\phi_0$=200 V). 
The input parameter fitted on the data calculated from PIC/MCC simulations (see Fig. \ref{FIGpotpic}) 
are used.}
\label{FIGtrajectory}
\end{center}
\end{figure}

Figure \ref{FIGtrajectory} shows the trajectories of dust particles calculated from Eq. (\ref{EQtrajectory}) and 
using the input parameters given above, for different values of the initial velocity. Right after the time of the 
abrupt phase shift all dust particles gain a certain initial velocity. If the initial velocity is below 
$u_0 \approx 1.0$ m/s, they cannot overcome the central maximum of the potential and bounce only inside the 
lower potential well. Dust particles with the initial velocity above $u_0 \approx 1.25$ m/s travel through the 
whole plasma bulk just after the phase shift. Dust particles with an initial velocity of $u_0 \approx 1.5$ m/s 
oscillate back and forth in the bulk region. However, their final equilibrium position is again located around 
the lower sheath. Therefore, from the model the initial velocity to realize the sheath-to-sheath transport is 
found at certain intervals, e.g., dust particles having $u_0$ = 2.0 m/s end up in the upper potential 
minimum while those having $u_0$ = 1.75 m/s do not. The conclusion obtained from Fig. \ref{FIGtrajectory} 
can be summarized by introducing the number of passages of dust particles through the plasma bulk, $N_{trans}$. 
\begin{table}
\begin{center}
\caption{\label{table1} Summary of the effective transport of dust particles through the plasma bulk 
obtained in this study, depending on the initial velocity (Ar, 8 Pa, $\phi_0=200$ V). Odd number of $N_{trans}$ realizes 
sheath-to-sheath transport, while even number of $N_{trans}$ does not.} 
\begin{tabular}{@{}|l|lllll|}
\hline
$u_0$ (m/s) & 1.00 & 1.25 & 1.5 & 1.75 & 2.0 \\
\hline
$N_{trans}$ & 0  &  1  &  2  &  2   &  3  \\
\hline
\end{tabular}
\end{center}
\end{table}
Any odd number of $N_{trans}$ means that sheath-to-sheath transport is realized, whereas even numbers of 
$N_{trans}$ correspond to a final position close to the initial position at the lower sheath edge (table \ref{table1}). 
We also note that the trajectory of $u_0 \approx$ 1.25 m/s 
obtained from the model agrees well with the experimental result (Fig. \ref{FIGtransport}(a)). 
\begin{figure}[t]
\begin{center}
\includegraphics{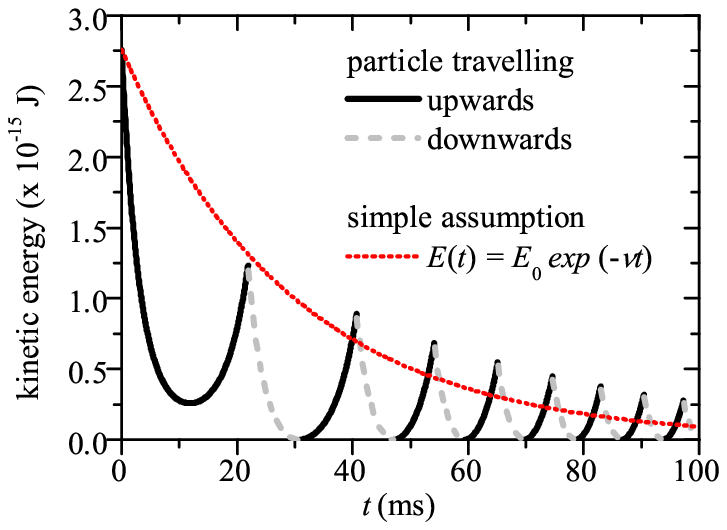}
\caption{Time evolution of the kinetic energy of dust particles after the abrupt phase shift according to 
the $u_0 = 1.25$ m/s case in Fig. \ref{FIGtrajectory} (Ar, 8 Pa, $\phi_0$ = 200 V).}
\label{FIGenergy}
\end{center}
\end{figure}

\indent Using Eq. (\ref{EQenergy}) the time evolution of the kinetic energy of the dust particles after 
the abrupt phase shift is obtained as shown in Fig. \ref{FIGenergy}. 
An anharmonic oscillation is superimposed 
on the simple assumption of an exponential decay of the dust velocity (Eq. (\ref{EQenergysimple})) as a function 
of time. The sharp edges in this oscillations are due to the treatment of the electrostatic forces as hard walls. 
When the dust particles bounce between the sheath edges, they do not just lose their kinetic energy on long 
timescales, but they also gain kinetic energy temporarily 
due to the ion drag force while moving from the discharge center towards the sheaths. 
However, the kinetic energy stays below $W_0 e^{-\nu t}$ between $t=$ 0 and the time of trapping 
in one of the two potential wells. This is because the potential profile leads to a deceleration of the dust 
particles just after the abrupt phase change. Therefore, the dust particles spend even more time on their way to 
the upper sheath and undergo more collisions with the neutral gas, resulting in enhanced friction losses. 
The information on the trajectory and energy provided by the analytical model of dust transport is useful for the 
optimization of their transport: it can be understood that a monoenergetic initial distribution within one of the 
velocity intervals allowing sheath-to-sheath transport, e.g., $u_0 \approx 1.25$ m/s in the case discussed 
here, is favorable to transport as many particles as possible to the upper sheath. Moreover, the outcome of the model 
suggests that the rough estimation of the probability of successful particle transport, $P_{trans}$, given above 
might overestimate the fraction of particles residing at the upper sheath edge, because the energy loss on the 
way from the upper sheath to the potential peak is much smaller than the energy loss occuring on the way from 
the lower sheath to the peak. In general, this model only requires the peak ion density in the discharge center 
and the electron temperature as input parameters, which could be measured by other diagnostic methods. However, 
there is no simple access to apply such methods in our experimental setup. The upgrading of the experimental setup 
to obtain these key parameters is required for our further study.

\subsection{Classification of transport conditions}
\begin{figure}[b]
\begin{center}
\includegraphics{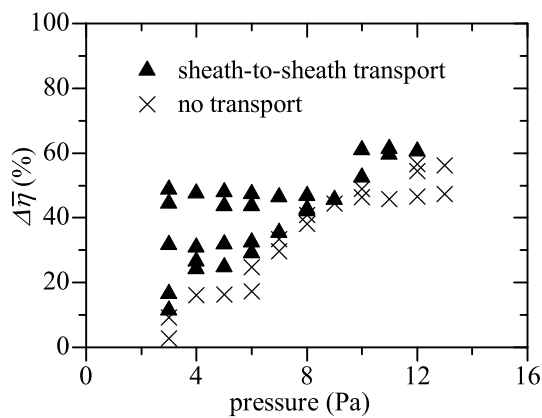}
\caption{Experimentally obtained classification of the dust particle transport as a function of $\Delta \bar{\eta}$ and pressure. 
The voltage amplitude is kept at $\phi_0$ = 200 V for $p<$10 Pa and $\phi_0$ = 200-240 V for $p\geq$10 Pa, respectively.}
\label{FIGtransclass}
\end{center}
\end{figure} 
\begin{figure}[ht]
\begin{center}
\includegraphics{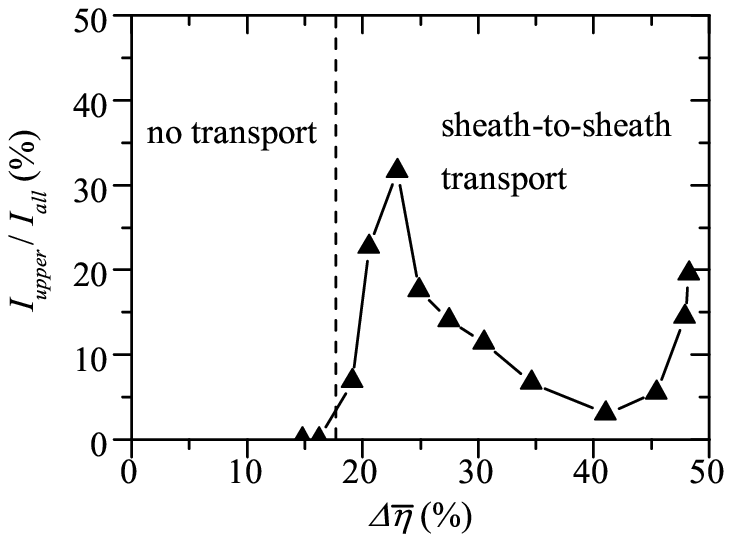}
\caption{Normalized measured LLS intensity from dust particles around the upper sheath edge 
($I_{upper} / I_{all}$) as a parameter of $\Delta \bar{\eta}$ for 
the abrupt phase shift (Ar, 4 Pa, $\phi_0$ = 200 V).
$I_{upper} / I_{all}$ is obtained by dividing the sum of the LLS intensity 
from dust particles around the upper sheath edge by that from dust particles around both sheath edges. }
\label{Velocity}
\end{center}
\end{figure}
We now turn to the discussion of conditions, under which sheath-to-sheath transport is possible. The key 
parameter for this transport is the rapid change of the dc self bias, $\Delta \eta$, which can be 
easily controlled between $\Delta {\eta}_{min}=0$ and $\Delta {\eta}_{max}= {\eta}(90^\circ) - {\eta}(0^\circ)$ 
by choosing certain intervals of the change in the phase angle (see Fig. \ref{FIGbias}).
As shown in Fig. \ref{FIGtransclass}, a threshold value of $\Delta \bar{\eta}$ is apparently required to 
achieve the transport of a fraction of the particles to the upper equilibrium position. 
Here the difference of normalized dc self bias $\Delta \bar{\eta}$ is given by 
$\Delta \bar{\eta} = [\eta(\theta_2) - \eta(\theta_1)] / \phi_0 $ in case of the phase shift 
from $\theta_1$ to $\theta_2$. The threshold increases with pressure, due to the increasing collisionality 
and, even more important, a stronger ion drag force, i.e. the central peak in the potential distribution becomes 
higher with increasing pressure. Therefore, it becomes more difficult for the particles to overcome this potential barrier. 
If $\Delta \bar{\eta}$ is smaller than the threshold, sheath-to-sheath transport is not realized: the dust particles 
reach a certain position below this potential peak and are forced towards the equilibrium position around the lower 
electrode sheath again (see Fig. \ref{FIGtransport}(b)). In this case, similar to the adiabatic phase change, 
information on the local plasma properties might be gained from this disturbance of the particle distribution. 
In particular, we observe that the maximum displacement of the dust particles strongly depends on 
global parameters, such as pressure and voltage, in the experiment.
However, a very good spatio-temporal resolution of the LLS measurements is required, which is not provided in our 
experiment. At low pressures, the sheath-to-sheath transport is possible within a wide range of $\Delta \bar{\eta}$ 
(see Fig. \ref{FIGtransclass}). However, as it has been motivated by the model results shown in Fig. \ref{FIGtrajectory}, 
the fraction of dust particles might vary as a function of $\Delta \bar{\eta}$. 
Figure \ref{Velocity} shows the normalized LLS intensity from dust particles around the upper sheath edge 
($I_{upper} / I_{all}$) as a function of $\Delta \bar{\eta}$, for the abrupt phase shift. A low pressure of 4 Pa has 
been applied here. $I_{upper} / I_{all}$ is obtained by dividing the sum of the LLS intensity from dust particles 
around the upper sheath edge by that from dust particles around both sheath edges. 
The maximum of $I_{upper} / I_{all}$ is seen at 
$\Delta \bar{\eta}$ = 23\%, and sheath-to-sheath transport is not achieved for $\Delta \bar{\eta}$ $<$ 16\%. 
These results indicate that the optimum initial velocity for sheath-to-sheath transport is slightly above the 
minimum value where sheath-to-sheath transport is realized. It also becomes clear that the change in the dc self 
bias, $\Delta \bar{\eta}$, for the {\it efficient} sheath-to-sheath transport is found at a certain interval, e.g., 
dust particles are transported efficiently for $\bar{\eta}$ = 48\% and $\bar{\eta}$ = 23\%, while they are not for 
$\bar{\eta}$ = 41\% (see Fig. \ref{Velocity}). The initial velocity of dust particles, $u_0$, is controlled by 
changing $\Delta \bar{\eta}$, since the temporally averaged sheath voltage depends almost linearly on the dc self 
bias \cite{EAEpower} and it can be approximated that the initial energy of the dust particles is proportional to 
the change of the mean sheath voltage. Hence, $u_0 \propto \sqrt{\Delta \bar{\eta}}$ and these results support 
the model of the dust motion described above (Fig. \ref{FIGtrajectory}). 
\\

\section{Conclusion}
The opportunities of controlling the transport of dust particles via the EAE have been discussed using the 
results of experiment, simulations, and analytical models. For these models, it has been confirmed that the 
dust particles do not significantly perturb the electrical properties of the discharge. In the case of an adiabatic 
tuning of the phase angle between the applied harmonics the dust particles are kept at an equilibrium position 
close to the lower sheath edge and their levitation is correlated with the time averaged electric field profile. 
This might provide the opportunity to estimate the electron density by using the dust particles as electrostatic probes. In 
the case of an abrupt phase shift ($90^{\circ}$ $\rightarrow$ $0^{\circ}$) the dust particles are transported upwards, 
i.e. they move between both sheaths through the plasma bulk. The trajectory as well as the temporal evolution of the 
dust particle energy are well understood using an analytical model. It is found that an initial velocity of the dust 
particles of about 1.25 m/s is required to push them over the potential hill located around the center of the plasma 
bulk. Thus, changing the applied voltage waveform via the EAE allows transporting a fraction of the dust particles from 
the equilibrium position around the lower sheath edge to the one at the upper electrode sheath, i.e. sheath-to-sheath 
transport is realized. The model also predicts that the initial velocity to realize sheath-to-sheath transport is 
found at certain intervals, which is in agreement with the dependence of the probability of sheath-to-sheath transport 
(fraction of LLS intensity at the upper sheath edge) on the change in the dc self bias found in the experiment. 
Furthermore, a certain threshold value of the rapid change of the dc self bias is 
required to achieve sheath-to-sheath transport. If the change in the dc self bias lies below the threshold value, 
the dust particles move within the lower potential well. 
Due to an increase in the collisionality and in the height of the potential peak, 
the threshold increases and the displacement decreases as a function of neutral gas pressure. 

\ack This research was supported by the German Federal Ministry for the Environment (0325210B), the 
Alexander von Humboldt Foundation, the RUB Research Department Plasma, and the Hungarian Scientific 
Research Fund (OTKA-K-77653+IN-85261, K-105476, NN-103150).

\end{document}